\documentclass[12pt]{article}
\usepackage{bm,amsmath,amssymb,graphicx}

\begin{document}
\pagenumbering{arabic}
\begin{titlepage}

\title{On the entropy of the massive conformal gravity universe}

\author{F. F. Faria$\,^{*}$ \\
Centro de Ci\^encias da Natureza, \\
Universidade Estadual do Piau\'i, \\ 
64002-150 Teresina, PI, Brazil}

\date{}
\maketitle

\begin{abstract}
We find that the total entropy of the massive conformal gravity universe is 
an increasing function of time, and therefore the cosmological model of the 
theory passes the generalized second law of thermodynamics test.
\end{abstract}

\thispagestyle{empty}
\vfill
\bigskip
\noindent * felfrafar@hotmail.com \par
\end{titlepage}
\newpage


\section{Introduction}
\label{sec1}


Although the standard $\Lambda$CDM cosmological model fits extremely well with 
several early and late universe data \cite{Ries,Agha}, it faces some important 
problems, such as the cosmological constant problem \cite{Wein}, the lithium 
problem \cite{Cybu}, the big bang singularity problem \cite{Haw1}, and the 
Hubble tension (see \cite{Verd} for a review), among others. There are a large 
number of theories that try to solve some of these problems but none of 
them have been completely successful so far. One of such theories is massive 
conformal gravity (MCG), which is a conformally invariant theory of 
gravity in which the gravitational action is the sum of the Weyl action 
with the Einstein-Hilbert action conformally coupled to a scalar field 
\cite{Far1}. 

So far it has been shown that the 
MCG cosmological model fits well with the Type Ia supernovae (SNIa) data 
without the cosmological constant problem \cite{Far2}, predicts the 
observed primordial abundances of light elements with a likely solution to 
the lithium problem \cite{Far3} and its observables are not harmed by
the big bang singularity \cite{Far4}. However, success in solving 
some problems found in the $\Lambda$CDM cosmological model is not enough 
to make the MCG cosmological model completely consistent with our universe. 
To achieve this, the model also has to pass other important cosmological 
tests. 

One of these tests is to check if the MCG cosmological model also fits 
well with overlapping SNIa, cosmic microwave background (CMB) and baryon 
acoustic oscillations (BAO) data without the Hubble tension. However, 
to check this  it is necessary the development of a theory for the growth 
of inhomogeneities in the model, what has not yet been done due to the 
complexity of the fourth-order MCG field equations. 

In this paper, we intend to do something simpler, but also important, by 
seeing if the MCG cosmology is consistent with the generalized second law of 
thermodynamics\footnote{Recent similar 
thermodynamic analyses in other alternative cosmological models can be seen 
in Ref. \cite{She,Sha,Ask,Sig,Yet} }. In Sec. 2, we describe the dynamics of 
the MCG universe. In Sec. 3, we study the validity of the generalized second 
law of thermodynamics in the MCG universe.  Finally, in Sec. 4, we present our 
conclusions.


\section{MCG universe}
\label{sec2}


The total action of MCG is given \cite{Far5}
\begin{eqnarray}
S &=& S_{\textrm{MCG}} + S_{m} \nonumber \\
&=&\int{d^{4}x} \, \sqrt{-g}\bigg[ \varphi^{2}R 
+ 6 \partial^{\mu}\varphi\partial_{\mu}\varphi 
- \frac{1}{2\alpha^2} C^2 \bigg] 
+ \frac{1}{c}\int{d^{4}x\mathcal{L}_{m}},
\label{1}
\end{eqnarray}
where $\varphi$ is a gravitational scalar field called 
dilaton, $\alpha$ is a coupling constant, $c$ is the speed of light in vacuum, 
\begin{equation}
C^2 = C^{\alpha\beta\mu\nu}C_{\alpha\beta\mu\nu} = R^{\alpha\beta\mu\nu}
R_{\alpha\beta\mu\nu} - 2R^{\mu\nu}R_{\mu\nu} + \frac{1}{3}R^{2}
\label{2}
\end{equation}
is the Weyl curvature invariant,
$R^{\alpha}\,\!\!_{\mu\beta\nu} 
= \partial_{\beta}\Gamma^{\alpha}_{\mu\nu} + \cdots$ is the Riemann tensor, 
$R_{\mu\nu} = R^{\alpha}\,\!\!_{\mu\alpha\nu}$ is the Ricci tensor, 
$R = g^{\mu\nu}R_{\mu\nu}$ is the scalar curvature, and 
$\mathcal{L}_{m} = \mathcal{L}_{m}[g_{\mu\nu},\Psi]$
is the conformally invariant Lagrangian density of the matter field 
$\Psi$.   

Varying the action (\ref{1}) with respect to $g^{\mu\nu}$ and $\varphi$, 
and noting that the dilaton field $\varphi$ acquires the spontaneously 
broken vacuum expectation value $\varphi_{0}$ below the Planck scale 
\cite{Mats}, we find the classical MCG field equations
\begin{equation}
\varphi_{0}^{2}G_{\mu\nu} - \alpha^{-2} B_{\mu\nu} 
= \frac{1}{2c}T_{\mu\nu},
\label{3}
\end{equation}
\begin{equation}
R = 0,
\label{4}
\end{equation}
where
\begin{equation}
G_{\mu\nu} = R_{\mu\nu} - \frac{1}{2}g_{\mu\nu}R
\label{5}
\end{equation}
is the Einstein tensor,
\begin{eqnarray}
B_{\mu\nu} &=& \Box R_{\mu\nu} 
- \frac{1}{3}\nabla_{\mu}\nabla_{\nu}R  -\frac{1}{6}g_{\mu\nu}\Box 
R + 2R^{\rho\sigma}R_{\mu\rho\nu\sigma} 
-\frac{1}{2}g_{\mu\nu}R^{\rho\sigma}R_{\rho\sigma}  \nonumber \\ &&
- \frac{2}{3}RR_{\mu\nu}  + \frac{1}{6}g_{\mu\nu}R^2
\label{6}
\end{eqnarray}
is the Bach tensor,
\begin{equation}
T_{\mu\nu} = - \frac{2}{\sqrt{-g}} \frac{\delta \mathcal{L}_{m}}
{\delta g^{\mu\nu}}
\label{7}
\end{equation}
is the matter energy-momentum tensor, and $\Box 
= \nabla^{\mu}\nabla_{\mu}$ . In addition, for $\varphi = \varphi_{0}$, the 
MCG line element $ds^2 = \left(\varphi/\varphi_{0}\right)^{2}g_{\mu\nu}dx^{\mu}dx^{\nu}$ 
reduces to  
\begin{equation}
ds^2 = g_{\mu\nu}dx^{\mu}dx^{\nu}.
\label{8}
\end{equation}
The full dynamics of the classical MCG universe can be described by
(\ref{3}), (\ref{4}) and (\ref{8})  without loss of generality.

The energy-momentum tensor of the dynamical perfect fluid that fills the MCG 
universe is given by \cite{Far2}
\begin{equation}
T_{\mu\nu} = 2cS_{0}^{2}\left(R_{\mu\nu} 
- \frac{1}{4}g_{\mu\nu}R\right) + \left( \rho + \frac{p}{c^2} \right)
u_{\mu}u_{\nu} + \frac{1}{4}g_{\mu\nu}\left( c^2\rho + p \right),
\label{9}
\end{equation}
where $c^2\rho$ is the energy density of the fluid, $p$ is the pressure of 
the fluid, $u^{\mu}$ is the
four-velocity of the fluid, which is normalized to $u^{\mu}u_{\mu} = - c^2$, 
and $S_0$ is a spontaneously broken constant vacuum expectation value of 
the Higgs field. 

By combining (\ref{9}) with (\ref{3}) and (\ref{4}), we find
\begin{equation}
\left(\varphi_{0}^{2}-S_{0}^{2}\right)R_{\mu\nu} 
- \alpha^{-2}B_{\mu\nu} = \frac{1}{2c}\left[ \left( \rho 
+ \frac{p}{c^2} \right)u_{\mu}u_{\nu} + \frac{1}{4}g_{\mu\nu}
\left( c^2\rho + p \right) \right].
\label{10}
\end{equation}
Then, substituting the Friedmann–Lema\^itre–Robertson–Walker (FLRW) metric
\begin{equation}
ds^{2} = - c^2dt^{2} + a(t)^2\left( \frac{dr^{2}}{1-Kr^{2}} +r^{2}d\theta^{2} 
+ r^{2}\sin^{2}\theta d\phi^{2} \right),
\label{11}
\end{equation}
the fluid four-velocity $u^{\mu} = (c, 0, 0 ,0)$, and\footnote{This 
value of $\varphi_{0}$ is necessary for the theory to be consistent with solar 
system observations \cite{Far6}.} $\varphi_{0}^{2} 
= 3c^3/32\pi G \gg S_{0}^{2}$ into (\ref{10}), and making some algebra, we 
obtain the MCG cosmological equations
\begin{equation}
H^2 + \frac{Kc^2}{a^2}  
=  \frac{4\pi G}{3}\left(1 + w \right)\rho,
\label{12}
\end{equation}
\begin{equation}
\dot{H} - \frac{Kc^2}{a^2}
= - \frac{8\pi G}{3}\left(1 + w \right)\rho,
\label{13}
\end{equation}
and the MCG energy continuity equation\footnote{Since (\ref{14}) does not depend 
on $w$, it is valid for both matter and 
radiation.}
\begin{equation}
\dot{\rho} + 4H\rho = 0,
\label{14}
\end{equation}
where the dot denotes $d/dt$,  $a = a(t)$ is the 
scale factor, $K = -1$, 0 or 1 is the spatial curvature, $G$ is the 
gravitational constant, $H = \dot{a}/a$ is the Hubble parameter, and 
$w = p/(c^2\rho)$ is the equation-of-state parameter, which is $0$ for 
non-relativistic matter (dust), $1/3$ for relativistic matter (radiation) 
and $-1$ for vacuum energy (dark energy)\footnote{We can see from (\ref{12}) and 
(\ref{13}) that the dark energy does not contribute to the dynamic evolution 
of the MCG universe.}. 

The direct integration of (\ref{14}) gives\footnote{It follows from (\ref{15}) 
that both matter and radiation evolve at the same rate in the MCG universe.}
\begin{equation}
\rho = \rho_{0}
\left(\frac{a_{0}}{a}\right)^{4},
\label{15}
\end{equation}
where the subscript $0$ denotes values at the present time 
$t_{0}$. By substituting (\ref{15}) into (\ref{12}), we obtain
\begin{equation}
\dot{a}^2   
=  \frac{b^2}{4a^2} - Kc^2,
\label{16}
\end{equation}
where $b = \sqrt{16 \pi G(1+w) \rho_{0} a_{0}^4/3}$. The solution to 
(\ref{16}) is given by\footnote{The fact that the MCG field equations 
are of fourth order in derivatives can lead to instabilities in the 
classical solutions of the theory. However, it was explicitly shown 
in Ref. \cite{Far2} that the MCG cosmological solution (\ref{17}) is stable.}
\begin{equation}
a = \sqrt{bt - Kc^2t^2},
\label{17}
\end{equation}
which is valid in all epochs of the MCG universe.


\section{Generalized second law of thermodynamics}
\label{sec3}


The generalized second law (GSL) of thermodynamics  establishes that the total 
entropy of the universe $S_{\mathrm{tot}}$, which is the sum of the entropy of 
the universe apparent horizon $S_h$ and the entropy of the cosmological fluid 
inside the horizon $S_f$, 
should be a non-decreasing function of time. We can express the GSL in the form
\begin{equation}
\dot{S}_{\mathrm{tot}} = \dot{S}_h + \dot{S}_f \geq 0.
\label{18}
\end{equation}

The entropy of the MCG universe apparent horizon can be derived from 
the entropy \cite{Mit1}
\begin{equation}
S_h = \left(\frac{k_{B}c^3}{4G\hbar}\right)A_{h} 
- \frac{8\pi^2 k_B}{\hbar c}\int{H R_h^4 (c^2\rho_{e}+ p_{e}) dt}
\label{19}
\end{equation}
of the apparent horizon of a cosmological model with the modified FLRW equations 
\begin{equation}
H^2 + \frac{Kc^2}{a^2}
= \frac{8\pi G}{3}\left(\rho + \rho_{e} \right),
\label{20}
\end{equation}
\begin{equation}
\dot{H} - \frac{Kc^2}{a^2}
= - \frac{4\pi G}{c^2}\left[c^2\left(\rho + \rho_{e} \right) 
+ \left(p + p_{e}\right)\right],
\label{21}
\end{equation}
where $k_B$ is the Boltzmann constant, $\hbar$ is the reduced 
Planck constant, $A_h$ is the surface area of the sphere with an 
apparent horizon radius
\begin{equation}
R_{\mathrm{h}} = \frac{c}{\sqrt{H^2 + Kc^2/a^2}},
\label{22}
\end{equation}
$\rho_{e}$ is an effective mass density, and $p_{e}$ is an 
effective pressure. It is not difficult to see that (\ref{19}) reduces 
to the standard Bekenstein-Hawking entropy \cite{Bek,Haw2} in the case 
of the general relativity universe, in which $\rho_{e} = p_e = 0$.

By comparing (\ref{12}) and (\ref{13}) with 
(\ref{20}) and (\ref{21}), we find that
\begin{equation}
\rho_{e} = - \frac{1}{2}\left( 1- w\right)\rho, \ \ \ \ \ 
p_{e} = \frac{1}{6}\left( 1 - 5w \right)c^2\rho,
\label{23}
\end{equation}
for the MCG universe. Substituting 
$A_h = 4\pi R_{\mathrm{h}}^2$ and (\ref{23}) into (\ref{19}), we obtain
\begin{equation}
S_h = \left(\frac{\pi k_{B}c^3}{G\hbar}\right)R_{h}^2 
+ \frac{8\pi^2 k_B c}{3\hbar}\int{H R_h^4 (1 + w)\rho \, dt}.
\label{24}
\end{equation}
The differentiation of (\ref{24}) with respect to time then gives
\begin{equation}
\dot{S}_h = \left(\frac{2\pi k_{B}c^3}{G\hbar}\right)R_{h}
\dot{R}_h + \left(\frac{8\pi^2 k_B c}{3\hbar}\right)H R_h^4 (1 + w)\rho.
\label{25}
\end{equation}
Using (\ref{13}) in the time derivative of (\ref{22}), we arrive at
\begin{equation}
\left(1 + w \right)\rho = \left(\frac{3 c^2}{8\pi G}\right)
\frac{\dot{R}_h}{H R_h^3}.
\label{26}
\end{equation}
Finally, the substitution of (\ref{26}) into (\ref{25}) leads to
\begin{equation}
\dot{S}_h = \left(\frac{3\pi k_{B}c^3}{G\hbar}\right)R_{h}\dot{R}_h,
\label{27}
\end{equation}
which is the rate of change of the entropy of the MCG universe apparent horizon.

In order to obtain the rate of change of the entropy of the cosmological fluid 
inside the MCG universe apparent horizon we use the Gibb's law \cite{Izq}
\begin{equation}
T_f dS_f = dE_f + pdV_h,
\label{28}
\end{equation}
where $T_f$ is the temperature of the fluid, $S_f$ is the entropy of the fluid,
\begin{equation}
E_f = (c^2\rho) V_h
\label{29}
\end{equation}
is the total energy of the fluid, and
\begin{equation}
V_h = \frac{4\pi}{3}R_h^3
\label{30}
\end{equation}
is the volume bounded by the horizon. By using (\ref{29}) in (\ref{28}), 
and doing some simple calculation, we obtain
\begin{equation}
\dot{S}_f = \frac{c^2}{T_f}\left[V_h\dot{\rho} 
+ \dot{V}_h (1 + w)\rho \right],
\label{31}
\end{equation}
which is valid for any cosmological model. 

It is usually assumed that $T_f$ is equal to the temperature 
of the apparent horizon $T_h$. However, we cannot assume this here 
because according to the Wien's law the thermal equilibrium between 
radiation, which always dominates the MCG universe \cite{Far3}, and 
the apparent horizon is impossible \cite{Min}. Instead, we consider 
that\footnote{It is worth noting that the Unruh temperature is positive 
for the MCG universe because its expansion is decelerated \cite{Far2}. 
In addition, since the MCG universe is always dominated by radiation, 
there is no problem in considering the Gibbons-Hawking temperature, which 
was obtained  for the static event horizon of the de Sitter space, for 
its apparent horizon.} 
\begin{equation}
T_f = T_u = \left( - \frac{\ddot{a}}{a}R_h^2\right)T_h 
= \left(\frac{\hbar c}{2 \pi k_b}\right)
\left(- \frac{\ddot{a}}{a}R_h\right),
\label{32}
\end{equation}
where $T_u$ is the Unruh temperature  experienced by a radial comoving 
observer infinitesimally close to the horizon \cite{Mit2} and 
$T_h = (\hbar c/2 \pi k_b) R_h^{-1}$ is the Gibbons-Hawking temperature 
of the apparent horizon \cite{Gib}. 

Using (\ref{12}), (\ref{13}) and (\ref{26}), we can write (\ref{32}) in the form
\begin{equation}
T_f = \left(\frac{\hbar c}{4 \pi k_b}\right)\frac{\dot{R}_h}{H R_h^2}.
\label{33}
\end{equation}
The substitution of (\ref{12}), (\ref{14}), (\ref{26}), (\ref{30}) and 
(\ref{33}) into (\ref{31}) then gives the rate of change of the entropy of 
the cosmological fluid inside the MCG universe apparent horizon
\begin{equation}
\dot{S}_f =  \left(\frac{3\pi k_{B}c^3}{G\hbar}\right)R_{h}\dot{R}_h,
\label{34}
\end{equation}
where we considered that the MCG universe is always dominated by 
radiation, i.e., $w = 1/3$.

It follows from (\ref{27}) and (\ref{34}) that
\begin{equation}
\dot{S}_{\mathrm{tot}} = \dot{S}_h + \dot{S}_f
= \left(\frac{6\pi k_{B}c^3}{G\hbar}\right)R_{h}\dot{R}_h,
\label{35}
\end{equation}
which is the rate of change of the total entropy of the MCG universe.
Finally, substituting (\ref{17}) and (\ref{22}) into (\ref{35}), 
we obtain
\begin{equation}
\dot{S}_{\mathrm{tot}} = \left(\frac{24 \pi k_{B}c^3}{G\hbar}\right)
\frac{c^2t(b-Kc^2t)(b-2Kc^2t)}{b^2}.
\label{36}
\end{equation}
As demonstrated in Figure $1$, the closed ($K=1$) MCG 
universe violates the GSL, since $\dot{S}_{\mathrm{tot}}$ 
becomes negative at certain values of $t$, indicating a decrease in 
entropy. On the other hand, $\dot{S}_{\mathrm{tot}}$ is always positive 
for the open ($K=-1$) and the flat ($K=0$) MCG universes, which means that 
these universes do not violate the GSL.\footnote{Although the 
flat MCG universe obeys the GSL, it is not consistent with the available 
observational values of $H_0$ and $t_0$\cite{Far2}, so that the open MCG 
universe   is the only physical cosmological solution to the MCG field 
equations.} 

\begin{figure}[h]
 \centering
	\includegraphics[scale=0.5]{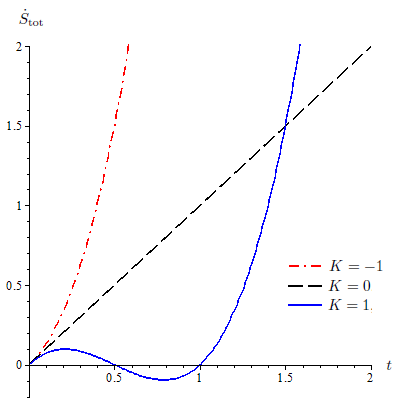}
	\caption{Rate of change of the total entropy of 
	the MCG universe versus the cosmic time for $K=-1$ (red dot-dashed), 
	$K=0$ (black dashed line) and $K=1$ (blue solid line). 
	Here, we assume $24 \pi k_{B}/G\hbar=b=c=1$, for simplicity.}
	\label{f2}
\end{figure}

Just to finish, we differentiate (\ref{36}) with respect to time, 
and assume that the MCG universe is open ($K = -1$), which gives
\begin{equation}
\ddot{S}_{\mathrm{tot}} = \left(\frac{24 \pi k_{B}c^3}{G\hbar}\right)
\frac{c^2(6c^4t^2 + 6bc^2t + b^2)}{b^2} > 0.
\label{37}
\end{equation}
This result leads us to conclude that the fluid and the horizon will never be in 
thermal equilibrium with each other in the open MCG universe, 
as expected.


\section{Final remarks}
\label{sec4}


Here, we have subjected the MCG cosmological model to the GSL test. To this aim,
we first constructed the rate of change of both the entropy of the MCG universe 
apparent horizon and the entropy of the MCG cosmological fluid inside the horizon. 
Then we added these two rates of change to obtain the rate of change of the 
total entropy of the MCG universe. The result found was that such rate of change 
is positive only for the open and flat MCG universes and, therefore, in these cases
the MCG universe pass the GLS test. In addition, we showed that the thermal 
equilibrium between the apparent horizon of the open MCG universe, which is 
the only one that is consistent with cosmological observations, and the MCG 
cosmological fluid inside the horizon will never be achieved, since the second time 
derivative of total entropy of the open MCG universe is also positive. 
This was expected once the MCG universe is radiation dominated in all epochs.

Although the result obtained here strengthens the credibility of the MCG 
cosmological model, a long way still has to be covered to prove that the model 
is fully consistent with the observed universe. Despite the great challenge, in 
future works, we intend to develop a theory for the grow of inhomogeneities in 
the model, which will open an opportunity for the study of the predictions of 
the model about the CMB and the BAO.



\begin{thebibliography}{99}

\bibitem{Ries}
A. G. Riess et al., Astron. J. \textbf{116}, 1009 (1998), arXiv:9805201; 
S. Perlmutter et al., ApJ \textbf{517}, 565 (1999), arXiv:9812133.

\bibitem{Agha}
N. Aghanim et al. [Planck Collab.], Planck 2018 results. VI. Cosmological parameters, 
Astron. Astrophys. \textbf{641}, A6 (2020), arXiv:1807.06209; 
Astron. Astrophys. \textbf{652}, C4 (2021) [erratum].

\bibitem{Wein}
S. Weinberg, Rev. Mod. Phys. \textbf{61}, 1 (1989).

\bibitem{Cybu}
R. H. Cyburt, B. D. Fields, K. A. Olive and T.-H. Yeh, Rev. Mod. Phys. \textbf{88}, 
015004 (2016), arxiv:1505.01076.

\bibitem{Haw1}
S. W. Hawking and G. F. R. Ellis, \textit{The large scale structure of space-time} 
(Cambridge University Press, England, 1973).

\bibitem{Verd}
L. Verde, T. Treu and A. G. Riess, 2019, Nature Astronomy \textbf{3}, 891 (2019), 
arXiv:1907.10625.

\bibitem{Far1}
F. F. Faria, Adv. High Energy Phys. \textbf{2014}, 520259 (2014), arXiv:1312.5553.

\bibitem{Far2}
F. F. Faria, Mod. Phys. Lett. A. \textbf{36}, 2150115 (2021), arXiv:1410.5104.

\bibitem{Far3}
F. F. Faria, Eur. Phys. J. C \textbf{83}, 81 (2023), arXiv:2301.11954.

\bibitem{Far4}
F. F. Faria,  Chin. J. Phys. \textbf{95}, 140 (2025), arXiv:2309.06389.

\bibitem{She}
A. Sheykhi, B. Wang and R.-G. Cai, Phys. Rev. D \textbf{76}, 023515 (2007), arXiv:0701261.

\bibitem{Sha}
M. Sharif and M. Zubair, JCAP \textbf{03}, 028 (2012), arXiv:1204.0848.

\bibitem{Ask}
M. Askin, M. Salti and O. Aydogdu, Mod. Phys. Lett. A. \textbf{32}, 1750177 (2017).

\bibitem{Sig}
O. Siginc, M. Salti, H. Yanar and O. Aydogdu, Mod. Phys. Lett. A. \textbf{24}, 1850137 (2018).

\bibitem{Yet}
U. Yeter, K. Sogut and M. Salti, Foundations of Physics \textbf{51}, 19 (2021).

\bibitem{Far5}
F. F. Faria, Adv. High Energy Phys. \textbf{2019}, 7013012 (2019), arXiv:1703.01318.

\bibitem{Mats}
N. Matsuo, Gen. Relativ. Gravit. \textbf{22}, 561 (1990).

\bibitem{Far6}
F. F. Faria, Mod. Phys. Lett. A. \textbf{37}, 2250033 (2022), arXiv:1604.02210.

\bibitem{Mit1}
S. Mitra, S. Saha and S. Chakraborty, Phys. Lett. B 734, 173 (2014), arXiv:1503.03059.

\bibitem{Bek}
J. D. Bekenstein, Phys. Rev. D \textbf{7}, 2333 (1973).

\bibitem{Haw2}
S. W. Hawking, Commun. Math. Phys. \textbf{43}, 199 (1975).

\bibitem{Izq}
G. Izquierdo and D. Pavón, Phys. Lett. B \textbf{633}, 420 (2006), arXiv:0505601.

\bibitem{Min}
J. P. Mimoso and D. Pavón, Phys. Rev. D \textbf{94}, 103507 (2016), 
arXiv:1610.07788.
 
\bibitem{Mit2}
S. Mitra, S. Saha and S. Chakraborty, Mod. Phys. Lett. A \textbf{30}, 1550058 
(2015), arXiv:1610.08050.

\bibitem{Gib}
G. W. Gibbons and S. W. Hawking, Phys. Rev. D \textbf{15}, 2738 (1977).


\end{thebibliography}
\end{document}